\documentclass[twocolumn,preprintnumbers,amsmath,amssymb,superscriptaddress,longbibliography,prl]{revtex4-1}

\usepackage{graphicx}
\usepackage{bm}
\usepackage{dsfont}
\usepackage[usenames,dvipsnames]{xcolor}
\usepackage{pstricks}
\usepackage[tight]{subfigure}
\usepackage{verbatim}
\usepackage{units}
\usepackage{multirow}
\usepackage{enumitem}
\usepackage{mathrsfs}
\usepackage{leftidx}
\usepackage{xspace}
\usepackage{braket}
\usepackage{bbm}
\usepackage{cancel}
\usepackage{amsfonts,amssymb,amsmath}

\usepackage[a4paper]{hyperref}
\hypersetup{colorlinks=true,linktoc=all,linkcolor=blue,breaklinks=true,citecolor=blue,urlcolor=blue}
\usepackage[english]{babel}
\newcommand{\lef}{\mathrm{L}}
\newcommand{\ri}{\mathrm{R}}

\newcommand{\RI}{R_I}
\newcommand{\RJ}{R_J}
\newcommand{\muI}{\Delta\mu_I}
\newcommand{\muJ}{\Delta\mu_J}
\newcommand{\RJz}{R_J^0}
\newcommand{\Dz}{D^0}

\usepackage[normalem]{ulem}

\usepackage{hyperref}

\begin{document}
		
	\title{{General bounds on} electronic shot noise in the absence of currents}
	
	\author{Jakob Eriksson}
	\affiliation{Department of Microtechnology and Nanoscience (MC2), Chalmers University of Technology, S-412 96 G\"oteborg, Sweden}
	\affiliation{University of Gothenburg, S-412 96 G\"oteborg, Sweden}
	
	\author{Matteo Acciai}
	\affiliation{Department of Microtechnology and Nanoscience (MC2), Chalmers University of Technology, S-412 96 G\"oteborg, Sweden}
	
	\author{Ludovico Tesser}
	\affiliation{Department of Microtechnology and Nanoscience (MC2), Chalmers University of Technology, S-412 96 G\"oteborg, Sweden}
	
	\author{Janine Splettstoesser}
	\affiliation{Department of Microtechnology and Nanoscience (MC2), Chalmers University of Technology, S-412 96 G\"oteborg, Sweden}
	
	\date{\today}
	
	\begin{abstract}

{We  investigate the charge and heat electronic noise in a generic two-terminal mesoscopic conductor in the absence of the corresponding charge and heat currents. Despite these currents being zero, shot noise is generated in the system. We show that, irrespective of the conductor's details and the specific nonequilibrium conditions, the charge shot noise never exceeds its thermal counterpart, thus establishing a general bound. Such a bound does not exist in the case of heat noise, which reveals a fundamental difference between charge and heat transport under zero-current conditions.}

	\end{abstract}

	\maketitle

\textit{Introduction.---}
In mesoscopic devices, noise arises both at equilibrium, due to thermal excitations leading to thermal noise~\cite{Johnson1927,Nyquist1928}, and out of equilibrium, where partitioning of electrons at a scatterer generates shot noise~\cite{Schottky1918}. Shot noise measurements have proven extremely useful to gain insights on charge carriers~\cite{Blanter2000} and have led, for instance, to the experimental observation of fractional charges~\cite{dePicciotto1997,Saminadayar1997}. Very recently, a distinct type of shot noise has been measured: this so-called delta-$T$ noise is purely due to a temperature bias {but is distinct from thermal noise} and is finite despite the \textit{average} current vanishes~\cite{Lumbroso2018,Sivre2019,Larocque2020}. This previously overlooked source of {nonequilibrium} noise was first introduced for diffusive conductors~\cite{Sukhorukov1999} and is currently attracting a lot of attention, both from a theoretical~\cite{Zhitlukhina2020,Rech2020,Hasegawa2020} and from an experimental perspective~\cite{Lumbroso2018,Sivre2019,Larocque2020}.
Until now, the analysis of this phenomenon
has however mostly been restricted by the following constraints: (i) conductors are assumed to have energy-independent transmission probabilities, (ii) the nonequilibrium state is induced by a pure temperature bias, (iii) only charge-current shot noise has been considered. 

In this Letter, we generalize this broadly by lifting all three constraints and thereby provide concrete results for a wider class of noise phenomena, of which the delta-$T$ noise is a specific manifestation.
Allowing the conductor to have
an energy-dependent transmission probability, we open up the analysis for two important aspects. Finite shot noise arises at zero-average current under generic nonequilibrium conditions,
e.g., combinations of temperature and voltage biases important for thermoelectrics.
Moreover, not only the charge, but also other types of currents are subject to shot noise persisting at zero current under appropriate nonequilibrium conditions: as an example,
we study the heat shot noise in the absence of an average heat current. Importantly, we find a fundamental bound for the charge shot noise under the condition of zero current: it can never exceed the thermal noise, independently of the system details and the type of nonequilibrium conditions. In contrast, such a bound does not exist for the heat shot noise in the absence of a heat current.

The presented analysis of zero-current charge  and heat shot noise will also be particularly relevant for the topical field of quantum thermodynamics~\cite{Benenti2017,Whitney2018May,Binder2018}. In nonequilibrium thermal engines, average currents between source and working substance can vanish~\cite{Sanchez2019Nov,Deghi2020Jul,Hajiloo2020Oct,Ciliberto2020Nov}, while their noise remains relevant~\cite{Sanchez2013,Freitas2020Nov,Marchegiani2020Nov}. Also the minimization of noise as a performance goal of  small-scale heat engines has recently been highlighted~\cite{Seifert2018Aug,Pietzonka2018May,Kheradsoud2019Aug,Pal2020May}, and correlations of heat and charge currents have been linked to their efficiency~\cite{Crepieux2014,Crepieux2016}.

\begin{figure}[!b]
\centering
\includegraphics[width=\columnwidth]{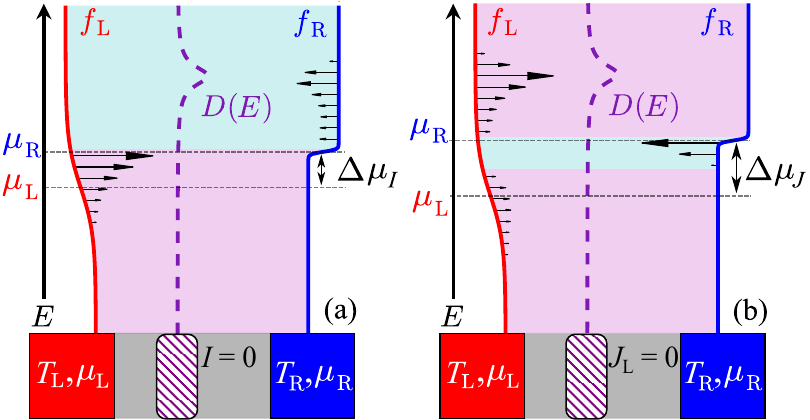}
\caption{Illustration of nonequilibrium conditions leading to vanishing charge~(a) and heat~(b) current into contact L of the setup sketched at the bottom. 
Fermi functions $f_\text{L}$ and $f_\text{R}$ of left and right contact with $T_\text{L}>T_\text{R}$ and with $\mu_\lef$ and $\mu_\ri$ are shown as red and blue lines. The purple dashed line shows the example of a Lorentzian transmission probability of the scatterer inside the conductor (gray). In order to fulfill the zero-current condition, we have (a)~$\mu_\ri-\mu_\lef=\muI$ and (b)~$\mu_\ri-\mu_\lef=\muJ$. Arrows show the magnitude and direction (emphasized by the background color) of the resulting energy-resolved fluxes adding up to zero-average charge or heat current.
}
\label{fig:sketch}
\end{figure}

Here, we study a generic two-terminal conductor as sketched at the bottom of Fig.~\ref{fig:sketch}, {using scattering theory}. The energy dependence of the conductor's transmission probability goes along with a combination of $\Delta \mu$ and $\Delta T$ fulfilling the requirement of a vanishing current. Considering a temperature bias $\Delta T$, the voltage required for zero charge current is the thermovoltage $\Delta\mu=\Delta\mu_I$, which naturally develops in a thermoelectric system under open-circuit conditions. Alternatively, it is possible to completely suppress the heat current into one of the terminals, despite heat being generally generated at nonequilibrium. For a given temperature bias, this suppression always requires the presence of a non-vanishing heat thermovoltage $\muJ$. In both situations, we find the corresponding shot noise to persist; we refer to this as the \textit{zero-current shot noise} in the following. This zero-current shot noise arises because opposite energy-resolved currents are present in the system (see Fig.~\ref{fig:sketch}). We find analytical expressions for the zero-current charge and heat shot noise {that can maximally be reached} in conductors with constant transmission, {requiring} large temperature bias. {Importantly}, these {limits can be exceeded,} when the transmission is energy-dependent. Indeed, the zero-current charge shot noise can approach the corresponding thermal noise and the heat shot noise is not bounded at all.

\textit{Formalism and model.---}
We consider a quantum conductor connected to two reservoirs ($\alpha=\lef,\ri$) characterized by Fermi distributions $f_\alpha(E)=\{1+\exp[\beta_\alpha(E-\mu_\alpha)]\}^{-1}$, where $\mu_\alpha$ are the chemical potentials and $\beta_\alpha=(k_\text{B}T_\alpha)^{-1}$ the inverse temperatures. In the following, we fix $\mu_\lef\equiv0$.
{We use} the framework of scattering theory~\cite{Blanter2000,Moskalets2011Sep}, where the conductor is described by a transmission probability, $D(E)$, that an electron at energy $E$ is transmitted from one reservoir to the other. {This description is valid for conductors with negligible or weak electron-electron interactions that can be treated at mean-field level. In} the recent measurements of delta-$T$ noise, scattering theory showed a very good agreement with experimental data in molecular junctions~\cite{Lumbroso2018}, quantum point contacts~\cite{Sivre2019} and tunnel junctions~\cite{Larocque2020}. 
In the following, we investigate the more general situation of zero-current charge and heat shot noise. Thus, we require vanishing charge current between the contacts, and heat current into contact $\alpha$. These are found from the expectation values of the operators  $\hat{X}=\hat{I}_\text{L}=-\hat{I}_\text{R}\equiv\hat{I}$ and  $\hat{X}=\hat{J}_{\alpha}=\hat{I}_\alpha^E-\mu_\alpha\hat{I}_\alpha$ and they are
\begin{equation}
X=\frac{2}{h}\int dE\,xD(E)[f_\text{L}(E)-f_\text{R}(E)]\,,
\label{eq:av-current}
\end{equation}
with $x\to\{-e,E-\mu_\alpha\}$ for $X\to\{I,J_\alpha\}$ and the elementary charge $e>0$.
The zero-frequency noise of these currents is defined as $S^X=2\int\braket{\delta \hat{X}(t)\delta \hat{X}(0)}dt$, where $\delta \hat{X}=\hat{X}-X$ is the fluctuation of the operator $\hat{X}$ around its average value $X$. The noise can be divided into two contributions $S^X=S_\text{th}^X+S_\text{sh}^X$,
\begin{subequations}
\label{eq:noise-general}
\begin{align}
    S_\text{th}^X&=\!\int\! dE \frac{4x^2}{h} D(E)\sum_{\alpha=\lef, \ri} f_\alpha(E)[1-f_\alpha(E)]\,,
    \label{eq:thermal-general}\\
    S_\text{sh}^X&=\!\int\! dE \frac{4x^2}{h} D(E)[1-D(E)][f_\lef(E)-f_\ri(E)]^2.
\label{eq:shot-general}
\end{align}    
\end{subequations}
The first term, $S_\text{th}^X$, is thermal-like noise. It contains independent contributions from the two contacts $\alpha=\lef, \ri$ and is therefore present even at equilibrium. The charge-current noise at equilibrium  is the famous Johnson-Nyquist noise~\cite{Johnson1927,Nyquist1928}. In contrast, the shot noise term $S_\text{sh}^X$ is only present \textit{out of equilibrium}, when $f_\lef\neq f_\ri$ and it is usually associated with the partitioning of a current flowing {through} the system. The standard situation is indeed when a pure voltage bias $\Delta\mu=\mu_\lef-\mu_\ri$ is applied, resulting in a net current $I\neq 0$. Then $S_\text{sh}^I$ reduces to conventional shot noise~\cite{Schottky1918}. However, as is clear from Eqs.~\eqref{eq:av-current} and \eqref{eq:shot-general}, a finite shot noise does not require a finite average current, $I\neq 0$. Indeed, another possibility to obtain a measurable shot noise, largely unexplored until recently, is to impose a pure temperature bias $\Delta T=T_\lef-T_\ri$. Then, it has been found for energy-independent transmission probabilities that $S_\text{sh}^I\neq 0$ even though $I=0$. This is the recently investigated delta-$T$ noise~\cite{Lumbroso2018,Sivre2019,Larocque2020}. 

{\textit{General bounds on charge and heat shot noise.---}}
We start by providing general bounds on the charge-current noise in the absence of a charge current for a generic two-terminal system, described by a transmission probability $D(E)$, and with an arbitrary voltage and temperature bias. By solving a variational problem following the method used in Ref.~\cite{Whitney2015Mar}, one can show that the uniform transmission probability $D(E) = 1/2$ maximizes the charge shot noise also under the condition of vanishing current~\cite{supp}. 
This uniform transmission leads at the same time to a possibly large thermal noise. Indeed, we find that for any transmission probability, the zero-current charge shot noise is bounded by the thermal noise. The inequalities
\begin{equation}
    S^I_\text{sh}\leq \frac{4e^2}{h}\int dE D(E) f_\text{L}(E)[1-f_\text{R}(E)] \leq S^I_\text{th},
    \label{eq:charge-bounds}
\end{equation}
can be obtained using only that $0\le \{D(E),f_\alpha(E)\}\le 1$, the monotonicity of $f_\alpha(E)$ and the zero-current condition. {Note that this inequality can be also extended to a system with many conduction channels and can hence even be applied to other types of conductors such as normal-superconducting junctions}~\cite{supp}. Hence, the charge \textit{shot} noise can never be greater than the \textit{thermal} noise, fixing the general bound
\begin{equation}\label{eq:bound}
    \RI\equiv\left[S_\text{sh}^I/S_\text{th}^I\right]_{I=0}\leq 1.
\end{equation}
Interestingly, at large temperature bias 
and with a small and sharp transmission probability, it is possible to approach equality in Eq.~(\ref{eq:bound}).
Concretely, this requires a transmission that is nonzero within an energy interval $\delta$ satisfying $\max(T_\text{L}, T_\text{R})\gg \delta/k_\text{B}\gg \min(T_\text{L}, T_\text{R})$ and $D(E)\ll 1$.
To show that such a transmission optimizes the noise ratio $R_I$, we assume without loss of generality that $T_\text{L}\gg T_\text{R}$. Then, the mentioned conditions allow us to approximate the shot noise $\frac{h}{4e^2}S^I_\text{sh}$ as
\begin{equation}
D^+f_\text{L}^2(\epsilon_0)(\epsilon_0+\frac{\delta}{2}-\mu_\text{R})+D^-(1-f_\text{L}(\epsilon_0))^2(\mu_\text{R}+\frac{\delta}{2}-\epsilon_0)\nonumber.
\end{equation}
Here, we call $\epsilon_0$ the energy around which $D(E)\neq0$ and
$D^\pm$ {is the integral mean value of $D(E)$} for $E\gtrless\mu_\text{R}$: {$D^\pm=(\epsilon_0-\mu_\text{R}\pm\delta/2)^{-1}\int_{\mu_\text{R}}^{\epsilon_0\pm\delta/2}D(E)\,dE$}. 
The zero-current condition fixes $\mu_\text{R}$ such that we obtain
\begin{equation}
    S^I_\text{sh}\approx \frac{4e^2}{h}\frac{\delta f_\text{L}(\epsilon_0)(1-f_\text{L}(\epsilon_0))D^+D^-}{D^-(1-f_\text{L}(\epsilon_0))+D^+f_\text{L}(\epsilon_0)}\approx S_\text{th}^I\,
\end{equation}
showing that indeed $R_I\to 1$. An example where this occurs is the Lorentzian function introduced in Eq.~(\ref{eq:lor}).

In contrast, the zero-current heat shot noise is \textit{not bounded} by the heat thermal noise, thus revealing a profound difference between charge transport and heat or energy transport.
{Essentially, this is due to the larger contribution to heat transport of electrons with higher energies. Choosing a transmission allowing for transport in two \textit{separate} energy regions and increasing the separation distance between them, the heat shot noise can become arbitrarily large~\cite{supp}, still maintaining a zero heat current and a limited thermal noise.
This does not apply to charge transport due to the electrons' fixed charge, thus limiting the magnitude of zero-current charge shot noise.}

\textit{Charge-current noise.---}
We now move on to show explicit results for the zero-current shot noise. First, we notice that if the temperatures satisfy $\Delta T\ll\bar{T}\equiv(T_\lef+T_\ri)/2$, the shot noise is very small compared with the thermal one (see for instance Ref.~\cite{Lumbroso2018}) both for charge and heat. Therefore, we focus henceforth on a large temperature bias, where the shot noise contribution can be significant. For a generic, weakly energy-dependent transmission function, we find for the ratio between the two noise contributions
\begin{equation}
    \RI=[1-D(0)]\left[\ln\left(2+2\cosh\frac{\Delta\mu_I}{k_\text{B}T_\lef}\right)-1\right]\,.
    \label{eq:ratio-charge}
\end{equation}
Here, $\beta_\text{L}\Delta\mu_I\propto D'(0) k_\text{B}T_\text{L}/\gamma$~\cite{supp} and $\gamma$ indicates a typical energy scale set by $D(E)$, while $D(E=\mu_\lef)=D(0)$ is the transmission in the vicinity of the transport window. This generalizes a previous result for delta-$T$ noise~\cite{Larocque2020} to arbitrary, weakly energy-dependent transmission probabilities, requiring only $\gamma\gg k_\text{B}T_\lef$. For a constant transmission, $D(E)=\Dz$, the thermovoltage is $\Delta\mu_I=0$ (as no thermoelectric effects arise in this particle-hole symmetric case), and Eq.~\eqref{eq:ratio-charge} reduces to
\begin{equation}
    \RI|_{D(E)=\Dz}=(1-\Dz)(2\ln{2}-1).
    \label{eq:charge-ratio-constant}
\end{equation}
Interestingly, a direct numerical evaluation of Eq.~\eqref{eq:noise-general} for  arbitrary temperatures $T_\lef$ and $T_\ri$~\cite{supp} shows that Eq.~\eqref{eq:charge-ratio-constant} is an \textit{upper} limit for $\RI$ in the case of a constant transmission.  
In contrast, we here find that a weak energy dependence in $D(E)$  already breaks this limit. In the regime of validity of Eq.~\eqref{eq:ratio-charge}, the presence of the thermovoltage $\Delta\mu_I\neq0$ always leads to an increase with respect to Eq.~\eqref{eq:charge-ratio-constant}.

We now compare the above generic features of a weakly energy-dependent transmission with the results we obtain for the specific case of a Lorentzian 
\begin{equation}
    D_\text{Lor}(E)=\Dz\frac{\Gamma^2}{\Gamma^2+(E-\epsilon_0)^2}\,,
    \label{eq:lor}
\end{equation}
with arbitrarily sharp energy dependence as indicated by the purple dashed line in Fig~\ref{fig:sketch}.
This experimentally relevant choice is of interest for highly efficient thermoelectrics~\cite{Hicks1993May,Mahan1996Jul,Benenti2017,Josefsson2018Oct} and is convenient  for our purposes since it can model a broad range of conductor characteristics. A resonant level with a small width $\Gamma$ is an example of a transmission with which the charge shot noise approaches the thermal noise, when $k_\text{B}\max(T_\text{L}, T_\text{R})\gg \Gamma\gg k_\text{B}\min(T_\text{L}, T_\text{R})$. In contrast, large $\Gamma\gg k_\text{B}\max(T_\text{L}, T_\text{R})$ reproduces the generic, weakly energy-dependent results of Eq.~(\ref{eq:ratio-charge}). The limit $\Gamma/ k_\text{B}\max(T_\text{L}, T_\text{R})\to\infty$, yielding a constant transmission, is the regime in which delta-$T$ noise was mainly studied previously; in the special case of $\Dz=1/2$ the shot noise is maximized.
\begin{figure}[t]
    \centering
    \includegraphics[width=\columnwidth]{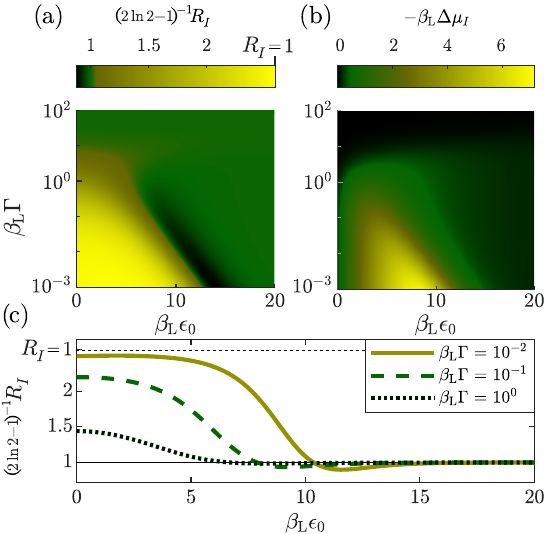}
    \caption{(a) Charge noise ratio $\RI$, normalized to the characteristic factor $2\ln 2-1$, see Eq.~(\ref{eq:charge-ratio-constant}). (b) Corresponding charge thermovoltage $\muI$. (c) Cuts of the density plot in (a) for different values of $\Gamma$. For all plots, we use $D(E)=D_\text{Lor}(E)$, with $\Dz=10^{-2}$, and $T_\ri\ll T_\lef$ at fixed, finite $T_\lef$. }
    \label{fig:c_plot}
\end{figure}

{Using Eq.~(\ref{eq:lor})}, we find $\Delta\mu_I$ by numerically solving the equation $I=0$.
Plugging $\Delta\mu_I$ into Eq.~\eqref{eq:noise-general} yields the resulting  zero-current charge shot noise and thermal noise. The charge noise ratio $\RI$, normalized to $2\ln 2-1$, together with the charge thermovoltage $\muI$ are shown in Fig.~\ref{fig:c_plot} as a function of the peak energy $\epsilon_0$  and the width $\Gamma$ of the Lorentzian, for $T_\lef\gg T_\ri$. Since $\muI$ $(R_I)$ is an odd (even) function of $\epsilon_0$, we consider $\epsilon_0\geq0$ only.

We find that the charge noise ratio {approaches} the value $\RI=2\ln 2-1$ (green areas) in those parameter regimes where $D_\text{Lor}(E)$ is close to constant. This {happens} in the limit $\beta_\lef\Gamma\gg 1$, where $D_\text{Lor}\to\Dz$, which we choose to be small compared with 1 ($\Dz=10^{-2}$, in Fig.~\ref{fig:c_plot}). {Likewise,} when $\beta_\text{L}\epsilon_0\gg1$, only the tail of the Lorentzian overlaps with the transport window (i.e.,~the region where the difference of Fermi functions in Eq.~\eqref{eq:av-current} is large) and the transmission function can be approximated by a constant much smaller than $\Dz$. Correspondingly, we find $\muI\to0$ in these regimes, see Fig.~\ref{fig:c_plot}(b).

Importantly, in those parameter regimes where the energy-dependence of the transmission probability cannot be neglected, the charge noise ratio $\RI$ exceeds the value $2\ln 2-1$ and approaches the fundamental bound $\RI=1$ [yellow regions in (b)]. This means that the zero-current charge shot noise can be equal to the thermal noise, as visible from Fig.~\ref{fig:c_plot}(c), showing that a sharp, weakly transmitting Lorentzian is an example of the optimal functions discussed after Eq.~(\ref{eq:bound}).
Interestingly, the limit $\RI\approx 1$ is not only reached at finite thermovoltages, but also for $\epsilon_0\rightarrow0$, where $\Delta\mu_I=0$ due to electron-hole symmetry. We conclude that even the charge noise ratio $\RI$ for delta-$T$ noise, namely for zero-current charge shot noise at $\muI=0$, can be increased by more than a factor 2 compared with the case of constant transmission.

\textit{Heat current noise.---}
 Importantly, the concept of zero-current shot noise is not limited to charge currents, but can greatly be extended to other transport quantities.  Here, as an example, we show explicit results for the zero-current heat shot noise, which to our knowledge has not been studied before. This choice is motivated by the current interest in mesoscopic heat engines, where fluctuations in heat and power can play an important role~\cite{Benenti2017}. Unlike conserved currents, the heat current in nonequilibrium conductors can in general be nullified only in one of the contacts at a time. We hence investigate 
 the heat current noise in contact L, $S^{J_\lef}\equiv S^J$, at vanishing heat current into the same contact.
 This situation of non-vanishing zero-current heat shot noise can only be obtained when both a temperature and a voltage difference across the conductor are present.

\begin{figure}[t]
    \centering
    \includegraphics[width=\columnwidth]{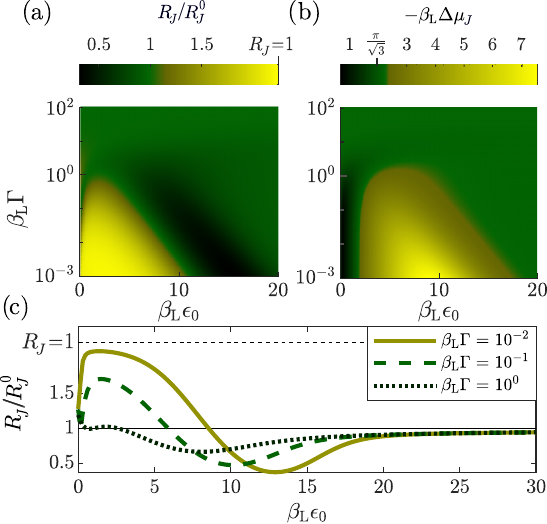}
    \caption{(a) Ratio $\RJ$ between the zero-current heat shot noise and heat thermal noise (normalized to the characteristic factor $\RJz$, see Eq.~(\ref{eq:heat-ratio-constant})). (b) Corresponding heat thermovoltage. (c) Cuts of the density plot in (a) for different values of $\Gamma$. For all plots, we use $D(E)=D_\text{Lor}(E)$, with $\Dz=10^{-2}$, and $T_\ri\ll T_\lef$ at fixed, finite $T_\lef$. }
    \label{fig:h_plot}
\end{figure}

Focusing again on the regime $T_\ri\ll T_\lef$ and considering first a generic, weakly energy-dependent transmission probability, with $k_\text{B}T_\lef\ll\gamma$, we find 
\begin{equation}
    \RJ=\frac{3}{\pi^2}[1-D(0)]A(x_J)\,,
    \label{eq:heat-ratio-general}
\end{equation}
where $x_J=\beta_\text{L}\Delta\mu_J$ is the dimensionless heat thermovoltage, $A(x)=2x^2\ln(1+e^x)-(\pi^2+x^3)/3+4x\text{Li}_2(-e^x)-4\text{Li}_3(-e^x)$, and $\text{Li}_n$ the polylogarithm function.
It is instructive to analyze Eq.~\eqref{eq:heat-ratio-general} when $D(E)=\Dz$. In this case, $x_J=\pm\pi/\sqrt{3}$~\footnote{This simple form is understood by equating the Joule heating $\Delta\mu_J^2/2$ to the conduction part of the heat current $\pi^2k_\text{B}^2T_\text{L}^2/6$} and 
we find
\begin{equation}
    \RJ|_{D(E)=\Dz}=\frac{3}{\pi^2}A\!\left(\frac{\pi}{\sqrt{3}}\right)(1-\Dz)\equiv \RJz(1-\Dz)\,.
    \label{eq:heat-ratio-constant}
\end{equation}
Since $\RJz\approx 0.45$, this shows that also the zero-current heat shot noise can be of the same order of magnitude as the corresponding thermal noise even for this simple transmission probability. As for the charge noise, we find that Eq.~\eqref{eq:heat-ratio-constant} provides an upper limit for $\RJ$ when $D(E)=\Dz$. This can already be exceeded for a weakly energy-dependent transmission, as $A(x_J)$ is an increasing function of $|x_J|$ and $x_J\propto\pm\pi/\sqrt{3}[1\pm\zeta k_\text{B}T_\text{L}/\gamma]$, where $\zeta$ is a coefficient depending on the transmission~\cite{supp}.

We finally move to the Lorentzian-shaped transmission function, going beyond the limit of a weak energy dependence. By solving $J_\lef=0$, we find $\Delta\mu_J$, which we plug into
Eq.~\eqref{eq:noise-general} to obtain the zero-current heat shot noise and heat thermal noise; their ratio for $T_\ri\ll T_\lef$ is plotted in Fig.~\ref{fig:h_plot}. 
The density plot for $\RJ$, displayed in Fig.~\ref{fig:h_plot}(a), shows extended regions (green) where the value $\RJ=\RJz$ is approached. As for the charge noise ratio, this occurs when $D(E)$ is close to constant and the heat thermovoltage [Fig.~\ref{fig:h_plot}(b)] approaches $-\beta_\lef\muJ=\pi/\sqrt{3}$.  
Interestingly, Fig.~\ref{fig:h_plot}(a) shows that $S_\text{sh}^J$ does not exceed $S_\text{th}^J$ even though the heat shot noise is generally unbounded. The reason for this is that the Lorentzian transmission probability allows transport of particles in one connected energy interval, only. Furthermore, Fig.~\ref{fig:h_plot} shows that, compared with $\RI$, there are more 
extended regions where $\RJ<R_J^0$.

\textit{Conclusions.---}
This work greatly extends previous studies on delta-$T$ noise to more general nonequilibrium conditions and to different types of transport quantities. 
We have proven a universal bound, namely that the zero-current charge shot noise can never exceed the zero-current thermal charge noise. For heat noise, such a bound does not exist.

While we have here focused on charge and heat transport, it would be interesting to investigate this concept for other kinds of currents such as spin currents and their noise in the future.
Moreover, an analysis of out-of-equilibrium fluctuation relations at vanishing average current~\cite{Altaner2016Oct,Maisel2019Dec,Safi2020} would be of interest for conductors studied here, not fulfilling local detailed balance conditions~\cite{Altaner2016Oct,Maisel2019Dec}.
Finally, the possible use of delta-$T$ noise for shot noise spectroscopy of thermal gradients has been pointed out~\cite{Lumbroso2018} and we expect this scope to become broader with extensions to different types of shot noise under general nonequilibrium conditions as presented in this Letter. 

\begin{acknowledgments}
We thank Rafael S\'anchez, Jens Schulenborg, Christian Sp\r{a}nsl\"{a}tt, and Robert Whitney for helpful comments on the manuscript. We acknowledge financial support from the Swedish VR (J.S.) and the Knut and Alice Wallenberg Foundation (M.A., J.S. and L.T.).
\end{acknowledgments}

\end{document}


\title{Supplemental material: General bounds on electronic shot noise in the absence of currents}
	
	\author{Jakob Eriksson}
	\affiliation{Department of Microtechnology and Nanoscience (MC2), Chalmers University of Technology, S-412 96 G\"oteborg, Sweden}
	\affiliation{University of Gothenburg, S-412 96 G\"oteborg, Sweden}
	
	\author{Matteo Acciai}
	\affiliation{Department of Microtechnology and Nanoscience (MC2), Chalmers University of Technology, S-412 96 G\"oteborg, Sweden}
	
	\author{Ludovico Tesser}
	\affiliation{Department of Microtechnology and Nanoscience (MC2), Chalmers University of Technology, S-412 96 G\"oteborg, Sweden}
	
	\author{Janine Splettstoesser}
	\affiliation{Department of Microtechnology and Nanoscience (MC2), Chalmers University of Technology, S-412 96 G\"oteborg, Sweden}
	
	\date{\today}
		\maketitle

{
\section{General bounds on charge and heat noises}

In this section, we provide detailed derivations of the bounds on the zero-current charge noises and the transmission probabilities which allow to reach these bounds. Moreover, an example of a transmission allowing for arbitrarily large zero-current heat shot noise is discussed in detail.

\subsection{Transmission probability maximizing the zero-current charge shot noise}
With the same strategy of Ref.~\cite{Whitney2015Mar} in mind, we use a variational approach to maximize the charge shot noise, while---importantly---keeping the charge current $I$ equal to zero.
Both shot noise and charge current are functionals of the transmission probability $D(E)$ and functions of the chemical potential difference. We choose $\mu_\text{L}=0$, and $\mu_\text{R}=\mu$ for simplicity.
First of all, we calculate the variation of the shot noise at a fixed temperature bias, which yields
\begin{equation}
    \delta S_\text{sh}^I =\frac{4e^2}{h} \int dE\left[ (1-2D)(f_\text{L}-f_\text{R})^2\delta D -D(1-D)2(f_\text{L}-f_\text{R})\frac{\partial f_\text{R}}{\partial \mu}\delta \mu\right].
\end{equation}
In this equation, both the variation $\delta D$ of the transmission and $\delta\mu$ of the chemical potential appear. Because of the zero current condition, these variations are not independent: indeed, by imposing that the current (at an arbitrary and given temperature bias) does not vary, we get
\begin{equation}\label{eq:deltaI=0}
    \delta I = -\frac{2e}{h}\left[\int dE(f_\text{L}-f_\text{R})\delta D - \int dE D\frac{\partial f_\text{R}}{\partial \mu}\delta\mu\right] =0 \implies \delta \mu = \left[\int dE(f_\text{L}-f_\text{R})\delta D \right]\left[ \int dE D\frac{\partial f_\text{R}}{\partial \mu}\right]^{-1}\,,
\end{equation}
thus expressing $\delta\mu$ in terms of $\delta D$. The interpretation of this relation is that, as a consequence of the variation in the transmission, there must be an appropriate modification of the chemical potential difference in order for the current to remain constant (and particularly equal to zero in our case).
Introducing the notation
\begin{equation}
    \phi = \int dED(1-D)2(f_\text{L}-f_\text{R})\frac{\partial f_\text{R}}{\partial \mu}, \quad \xi = \int dED \frac{\partial f_\text{R}}{\partial \mu},
\end{equation}
and using the relation obtained in Eq.~(\ref{eq:deltaI=0}), we find that the variation of the shot noise is
\begin{equation}
    \delta S_\text{sh}^I =\frac{4e^2}{h} \int dE(f_\text{L}-f_\text{R})^2 \delta D \left[1-\frac{\phi}{\xi}\frac1{f_\text{L}-f_\text{R}} - 2D \right]=\frac{8e^2}{h} \int dE(f_\text{L}-f_\text{R})^2 \delta D \left[\widetilde{D}-D\right].
\end{equation}
Since $\delta D$ is arbitrary and $f_\text{L}\neq f_\text{R}$ at nonequilibrium, it is the expression inside the square brackets that determines the transmission probability maximizing the shot noise. In fact, the function
\begin{equation}
    \widetilde{D}(E) = \frac12 -\frac{\phi}{\xi}\frac{1}{2(f_\text{L}-f_\text{R})}
\end{equation}
would yield the maximum shot noise but it is not a transmission probability because it can be negative or greater than one. Thus, the transmission probability that maximizes the shot noise is
\begin{equation}
    D(E) = \left\{\begin{array}{ll}
          0 & \text{if}\quad \widetilde{D}(E)< 0 \\
          \widetilde{D}(E)& \text{if}\quad 0\leq\widetilde{D}(E)\leq 1\\
          1 & \text{if}\quad\widetilde{D}(E)> 1 \ .
    \end{array}\right.
    \label{eq:supp:transmission-max}
\end{equation}
We note that, in principle, this is a complicated self-consistent integral equation for $D(E)$. Moreover, the obtained transmission function still depends on $\mu$, that appears in $\phi$, $\xi$ and $f_\text{R}$. This is because we have not imposed the zero-current condition yet. This can be done by explicitly calculating the charge current with the transmission function \eqref{eq:supp:transmission-max}, obtaining $I$ as a function of $\mu$ and then imposing $I(\mu)=0$ to determine it. The calculation is greatly simplified by assuming that $T_\text{R}$ is the smallest energy scale in the system, allowing us to approximate $\phi/\xi\rightarrow 2(1-D(\mu))(f_\text{L}(\mu)- f_\text{R}(\mu))$. Then, with this expression, one notices that
\begin{equation}
    \frac{\delta S_\text{sh}^I}{\delta D(\mu)}=-\frac{2e^2}{h}[f_\lef(\mu)-f_\ri(\mu)]^2\le 0\,.
\end{equation}
This means that, unless $\mu=0$, an increase of $D(\mu)$ reduces the shot noise. If $\mu=0$, one immediately finds $D(E)=1/2$ and also a vanishing current. If this is not the case, we have to choose the minimum possible value for the transmission at $E=\mu$, i.e.~$D(\mu)=0$. Then,
\begin{equation}
    \widetilde{D}(E)=\frac{1}{2}-\frac{f_\lef(\mu)-f_\ri(\mu)}{f_\lef(E)-f_\ri(E)}
\end{equation}
and we can explicitly calculate the current with this transmission function, bearing in mind the constraints provided by Eq.~\eqref{eq:supp:transmission-max}. We find an odd function $I(\mu)=-I(-\mu)$, where
\begin{equation}
    I=\frac{\hl{2}e}{h}k_\text{B}T_\lef
    \begin{cases}
    \frac{2\ln(e^{2\tilde{\mu}}-1)-4\ln 2-\tilde{\mu}-e^{\tilde{\mu}}[\tilde{\mu}-4\coth^{-1}(e^{\tilde{\mu}})]}{2(1+e^{\tilde{\mu}})} & 0<\tilde{\mu}<\ln 2\\
    \ln(1+e^{\tilde{\mu}})-\tilde{\mu}+\frac{\ln[(e^{\tilde{\mu}}-1)/4]-\tilde{\mu}+e^{\tilde{\mu}}[\tilde{\mu}-\ln(e^{\tilde{\mu}}-1)]}{2(1+e^{\tilde{\mu}})} & \tilde{\mu}>\ln 2
    \end{cases}
\end{equation}
and $\tilde{\mu}=\beta_\lef\mu$.
From this result, one finds that $\mu\to 0$ is needed to nullify the current, which then yields the solution to the problem: $D(E)=1/2$.
This is the standard result that would also be obtained by maximizing the shot noise at a given nonequilibrium condition (i.e.~at fixed voltage and temperature biases), but was not necessarily expected here, where the zero-current condition is a constraint in the optimization procedure.

\subsection{Bound on the zero-current charge noise ratio}\label{supp:bound}
In this section, we derive an inequality between shot and thermal charge current noise contributions under the condition of zero average charge current. No assumptions on the properties of the transmission probability or on the temperature or the potentials of the contacts is made here. Since in this section we deal with charge current noise only, for brevity, we here always write shot noise and thermal noise without referring to ``charge current".

From the definition of the shot noise, a first inequality arises from noticing that the function $[1-D(E)]$ takes values between zero and one,
\begin{equation}
    \frac{h}{4e^2}S^I_\text{sh} = \int dED[1-D] (f_\text{L}-f_\text{R})^2 \leq \int dED (f_\text{L}-f_\text{R})^2 = \int dED (f_\text{L}-f_\text{R})[(1-f_\text{R})-(1-f_\text{L}))].
\end{equation}
Here, the \textit{equality} is satisfied only when $D(E) = 0$ or $f_\text{L}=f_\text{R}$, namely when the shot noise is zero.
In the last expression, we recognize a contribution which equals the negative of the thermal noise.
Therefore, we can write the inequality as
\begin{equation}\label{eq:Ssh+Sth<}
    \frac{h}{4e^2}\left(S^I_\text{sh} + S^I_\text{th}\right) \leq \int dED[f_\text{L} + f_\text{R} -2f_\text{L}f_\text{R}] = 2\int dEDf_\text{L}(1 -f_\text{R}) = 2\int dEDf_\text{R}(1 -f_\text{L}),
\end{equation}
where the last equalities are obtained using the zero-current condition.
From the definition of the thermal noise, we can write
\begin{equation}\label{eq:Sth=Omega}
\frac{h}{4e^2}S^I_\text{th} = \int dED f_\text{L}(1-f_\text{R}) + \int dED[f_\text{R} - f_\text{L}^2 -f_\text{R}^2 + f_\text{L}f_\text{R}],
\end{equation}
where we recognize the first integral from the inequality of Eq.~\eqref{eq:Ssh+Sth<}. Now, we proceed by proving that the second integral of Eq.~\eqref{eq:Sth=Omega} is always non negative.
Exploiting the zero-current condition, and using that the Fermi distribution takes values between zero and one, we have
\begin{equation}\label{eq:Sth>Omega}
    \begin{split}
    \int dED\left[\frac{f_\text{L} + f_\text{R}}{2} - f_\text{L}^2 -f_\text{R}^2 + f_\text{L}f_\text{R}\right] &\geq \int dED\left[\frac{f_\text{L} + f_\text{R}}{2} - f_\text{L}^2 -f_\text{R}+ f_\text{L}f_\text{R}\right] \\
    & = \int dED\left[\frac{f_\text{L} - f_\text{R}}{2} - f_\text{L}^2+ f_\text{L}f_\text{R}\right] 
     = -\int dEDf_\text{L}[f_\text{L}-f_\text{R}].
    \end{split}
\end{equation}
Assuming that the left temperature is higher than the right one, $T_\text{L}>T_\text{R}$, and calling $\epsilon$ the energy for which $f_\text{L}(\epsilon) =f_\text{R}(\epsilon) $, we can split the integral based on the sign of the integrand. Exploiting the fact that the Fermi function is a decreasing function of energy we have
\begin{equation}\label{eq:int>I=0}
    \begin{split}
     -\int dEDf_\text{L}[f_\text{L}-f_\text{R}] &= \int_{-\infty}^{\epsilon} dEDf_\text{L}[f_\text{R}-f_\text{L}] - \left(\int_\epsilon^\infty Df_\text{L}[f_\text{L}-f_\text{R}]\right)\\
     &\geq f_\text{L}(\epsilon)\int_{-\infty}^{\epsilon} dED[f_\text{R}-f_\text{L}] - f_\text{L}(\epsilon)\int_\epsilon^\infty dED[f_\text{L}-f_\text{R}] = f_\text{L}(\epsilon)(-I) =0.
     \end{split}
\end{equation}
If the right temperature were higher than the left one, $T_\text{R}>T_\text{L}$, we would get to the same conclusion by swapping the subscripts L, R in Eqs.~\eqref{eq:Sth>Omega} and \eqref{eq:int>I=0}.
Following these inequalities we have a lower bound for the thermal noise, and combining it with the inequality of Eq.~\eqref{eq:Ssh+Sth<}, we also find an upper bound for the shot noise,
\begin{equation}\label{eq:Sth>Ssh}
    \frac{h}{4e^2}S^I_\text{th}\geq \int dED f_\text{L}[1-f_\text{R}] = \int dED f_\text{R}[1-f_\text{L}]\geq \frac{h}{4e^2}S^I_\text{sh},
\end{equation}
which is the result provided in the main text.
{\subsection{Generalization to a multi-channel conductor}
When the conductor connecting the reservoirs has $N$ transport channels, we can express the charge current and noise in the basis of eigen-channels as 
\begin{gather}
    I = -\frac{2e}{h} \int dE \sum_n D_n(f_\text{L}- f_\text{R}),\\
    S^I = S^I_\text{sh} + S^I_\text{th} = \frac{4e^2}{h}\int dE \sum_n \left[D_n(1-D_n)(f_\text{L}-f_\text{R})^2 + D_n f_\text{L}(1-f_\text{L}) + D_n f_\text{R}(1-f_\text{R})\right],
\end{gather}
where $D_n = D_n(E)$ is the transmission probability of the $n$-th eigen-channel.
Introducing an effective single-channel transmission probability $\widetilde{D}(E) = N^{-1}\sum_n D_n(E)$, we can write the current and the noises as
\begin{gather}
    I = -\frac{2e}{h} \int dE N\widetilde{D}(f_\text{L}- f_\text{R}) = N\widetilde{I}, \\
    S^I_\text{th} = \frac{4e^2}{h}\int dE N\widetilde{D}[f_\text{L}(1-f_\text{L}) + f_\text{R}(1-f_\text{R})] = N\widetilde{S}^I_\text{th}, \\
    S^I_\text{sh} = \frac{4e^2}{h}\int dE N \widetilde{D}(1-\widetilde{D})(f_\text{L}-f_\text{R})^2 +\frac{4e^2}{h}\int dE\left[N\widetilde{D}^2-\sum_n D_n^2\right](f_\text{L}-f_\text{R})^2 = N \widetilde{S}^I_\text{sh} +C,\label{eq:supp:multi-channel-shot}
\end{gather}
where $\widetilde{S}^I_{\text{th(sh)}}$ is the thermal (shot) noise generated by a single-channel conductor described by the effective transmission $\widetilde{D}$.
The integrand of the additional contribution to the shot noise $C$ is always non-positive due to the convexity of the square function, which yields
\begin{equation}
    N\widetilde{D}^2 =\frac1N \left(\sum_n D_n\right)^2 \leq \sum_n D_n^2,
\end{equation}
meaning that $C\leq0$.
Using the result obtained in Sec.~\ref{supp:bound} for a single-channel conductor with transmission $\widetilde{D}$, we know that in the zero-current condition $I = \widetilde{I} =0$, the inequality $\widetilde{S}^I_\text{th} \geq \widetilde{S}^I_\text{sh}$ holds. Combining it with Eq.~\eqref{eq:supp:multi-channel-shot}, we obtain the following chain of inequality
\begin{equation}
   S^I_\text{th}= N\widetilde{S}^I_\text{th} \geq N\widetilde{S}^I_\text{sh} \geq N\widetilde{S}^I_\text{sh} + C = S^I_\text{sh},
\end{equation}
which proves the validity of the bound also in the multi-channel case.}

Interestingly, thanks to this result, our central findings also apply to other types of conductors, such as normal-superconducting junctions. Indeed, in that case, quasiparticle transport can be addressed by a generalized scattering approach that still relies on a unitary scattering matrix and with the addition of a transport channels for holes. The presence of the superconducting gap is taken into account by a proper transmission function.

\subsection{Arbitrarily large zero-current heat shot noise}
\begin{figure}[t]
\includegraphics[width=0.3\textwidth]{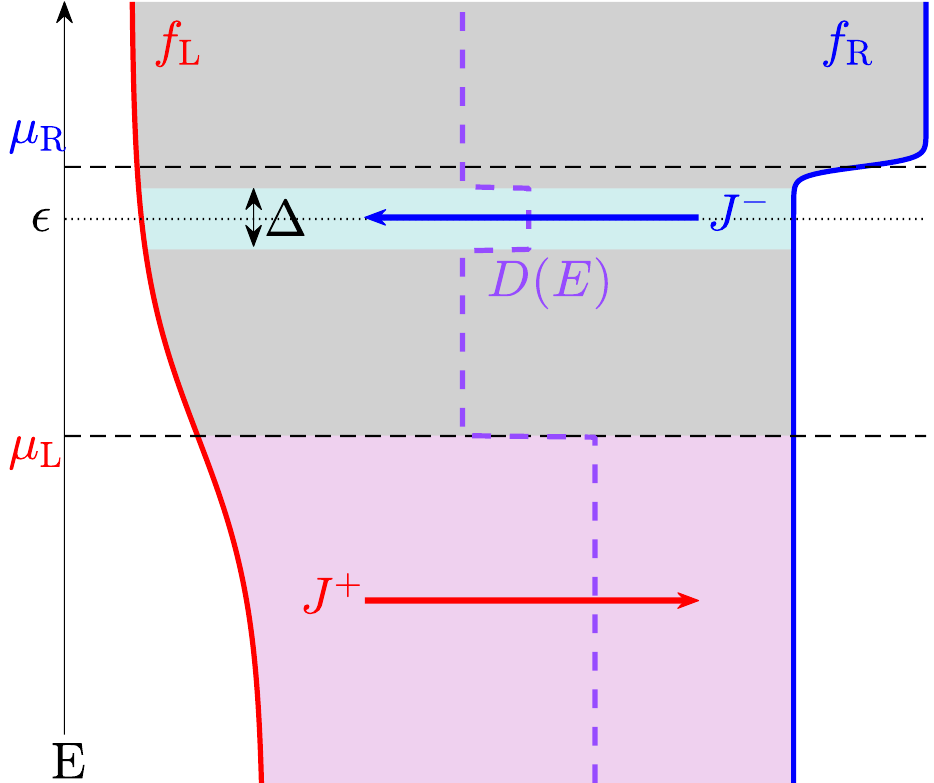}
\caption{Schematic representation of the energy landscape of the system under the condition of zero heat current, $J_\text{L}=0$. The transmission probability $D(E)$ (purple dashed line) allows for transport only for $E<\mu_\text{L}$ or  in the interval $[\epsilon-\Delta/2, \epsilon+\Delta/2]$. Transport at energies below $\mu_\text{L}$ yields a positive contribution to the heat current, $J^+$, which is balanced by the negative contribution, $J^-$, generated by the transport in the energy interval $[\epsilon-\Delta/2, \epsilon+\Delta/2]$.}
\label{fig:Dheatshot}
\end{figure}

To show that a bound similar to that in Eq.~\eqref{eq:Sth>Ssh} does not exist for the heat noises, it is sufficient  to provide an example in which the thermal noise is limited while the shot noise is not. This can be achieved with a transmission probability that leads to (1) a fixed amount of both thermal and shot noise in the energy interval in which the Fermi function of the hotter reservoir is smeared and (2) allows for heat transport  which can counterbalance the heat current in (1) and leads to a large heat shot noise but negligibly small thermal noise in the vicinity of the electrochemical potential of the colder reservoir.
With this purpose, we consider the transmission probability
\begin{equation}
    D(E) = \left\{\begin{array}{ccl}
         1 & \mathrm{for}&E<0 \\[0.25cm]
         0& \mathrm{for}&0\leq E < \epsilon -\frac\Delta{2}  \\[0.25cm]
         \frac12& \mathrm{for}& \epsilon -\frac\Delta{2}\leq E < \epsilon +\frac\Delta{2}\\[0.25cm]
         0 & \mathrm{for} &\epsilon +\frac\Delta{2}\leq E
    \end{array}\right.
\end{equation}
which allows for transport at all negative energies and in a small window of width $\Delta$ at positive energies. The width $\Delta$ will be determined by imposing the zero-current condition. We now choose to set the right chemical potential $\mu_\text{R}$ to be above the transmission window at positive energies, namely $\mu_\text{R}>\epsilon+\Delta/2$. We can then split the energy current in a positive contribution and a negative one
\begin{equation}
    J = \frac{2}{h}\int dE ED[f_\text{L}-f_\text{R}] = \frac{2}{h}\int_{-\infty}^0 dE E [f_\text{L}-f_\text{R}] + \frac{1}{h}\int_{\epsilon-\Delta/2}^{{\epsilon+\Delta/2}}dE E [f_\text{L}-f_\text{R}] = J^+ + J^-,
\end{equation}
as depicted in Fig.~\ref{fig:Dheatshot}.
The heat current $J^+$ is a finite and positive quantity, while the heat current $J^-$ is negative. When the right temperature is the smallest energy scale we can approximate $f_\text{R}$ with a step function. Moreover, when the position of the transmission window $\epsilon$ is much larger than the left temperature $T_\text{L}$ and the width $\Delta$, we can approximate the heat current as
\begin{equation}
    J = J^+ + J^- \approx J^+-\frac{1}{h} \epsilon \Delta .
\end{equation}
Imposing the condition of zero heat current, we get
\begin{equation}
    J  =0\Rightarrow \Delta =\frac{hJ^+}{\epsilon}.
\end{equation}
Now, we evaluate the heat noises. We notice that only the transmission window at positive energies contributes to the heat shot noise for the chosen transmission probability, since otherwise we always have $D=0$ or $D=1$. Thus, we find
\begin{equation}
    S^J_\text{sh} = \frac{1}{h}\int_{\epsilon-\Delta/2}^{{\epsilon+\Delta/2}}dE E^2 [f_\text{L}-f_\text{R}]^2 \approx \frac{\epsilon^2\Delta}{h} = \epsilon J^+.
\end{equation}
Since $J^+$ does not depend on the position of the positive-energy transmission window, $\epsilon$, we notice that the heat shot noise increases linearly with $\epsilon$. Therefore, the heat shot noise is unbounded from above.
Under the same assumptions, we notice that the positive-energy transmission window contributes weakly to the heat thermal noise because the left Fermi distribution $f_\text{L}$ decreases exponentially, and we can approximate the thermal noise as
\begin{equation}
    S^J_\text{th} \approx \frac{4}{h}\int_{-\infty}^0 dE E^2 f_\text{L}[1-f_\text{L}],
\end{equation}
which does not depend on $\epsilon$. Therefore, for a sufficiently large $\epsilon$, the heat shot noise is greater than the thermal noise.
}
\section{Charge noise for a general transmission}
\label{sec:app:charge-large}
In this section, we evaluate the charge shot noise at the thermovoltage for a general transmission function $D(E/\gamma)$. Here, $\gamma$ is a typical energy scale in the system. 

We are interested in evaluating the shot noise in the limit where one of the two temperatures is much smaller than the other; without loss of generality, we then take $T_\lef=T$ and $T_\ri=0$. We also consider $\mu_\lef=0$ and $\mu_\ri=-\Delta\mu$, where $\Delta\mu$ is an up to now arbitrary voltage bias. Later on, we will set $\Delta\mu=\Delta\mu_I$, where $\Delta\mu_I$ is the thermovoltage which develops in the system in response to the temperature bias in such a way that the charge current vanishes. With the above assumptions, and defining the function $F(E/\gamma)=D(E/\gamma)[1-D(E/\gamma)]$, the charge shot noise reads
\begin{equation}
\begin{split}
    S^I_{\text{sh}}&=\frac{4e^2}{h}\int_{-\infty}^{+\infty}dE\,F(E/\gamma)[f_\lef(E)-f_\ri(E)]^2\\
    &=\frac{4e^2}{h}\int_{-\infty}^{+\infty}dE\,F(E/\gamma)\left[f_E(E)+k_\text{B}Tf_\lef'(E)-2f_\lef(E)\Theta(-E-\Delta\mu)+\Theta(-E-\Delta\mu)\right]\\
    &=\frac{4e^2}{h}\left[\int_{-\Delta\mu}^{+\infty}dE\,F(E/\gamma)f_\lef(E)+\int_{\Delta\mu}^{+\infty}dE\,F(-E/\gamma)f_\lef(E)+k_\text{B}T\int_{-\infty}^{+\infty}dE\,F(E/\gamma)f'_\lef(E)\right]
\end{split}
\label{eq:charge-shot-general}
\end{equation}
where $\Theta(x)$ is the Heaviside step function. In order to proceed analytically, we assume that the temperature $T_\lef\equiv T$ in the left reservoir is small compared to the typical energy scale $\gamma$ determined by the transmission function. In other words, this physically means that the transmission function is weakly energy-dependent on the scale of the relevant transport window. To summarize, the approximations we are considering are $k_\text{B}T_\ri\ll k_\text{B}T_\lef\ll\gamma$.

Consider now the integral
\begin{equation}
    \mathcal{I}=\int_{-\Delta\mu}^{+\infty}dE\,F(E/\gamma)f_\lef(E)=k_\text{B}T\int_{-\frac{\Delta\mu}{k_\text{B}T}}^{+\infty}d\omega\,F\left(\frac{k_\text{B}T}{\gamma}\omega\right)\frac{1}{1+e^{\omega}}\,.
\end{equation}
For $k_\text{B}T\ll\gamma$, it can be approximated as
\begin{equation}
\begin{split}
    \mathcal{I}&=k_\text{B}T\left\{F(0)\left[x+\ln\left(1+e^{-x}\right)\right]+\frac{k_\text{B}T}{\gamma}F'(0)\left[-x\ln\left(1+e^{x}\right)-\text{Li}_2\left(-e^{x}\right)\right]\right.\\
    &\quad\left.+\frac{F''(0)}{2}\left(\frac{k_\text{B}T}{\gamma}\right)^2\left[x^2\ln\left(1+e^{x}\right)+2x\text{Li}_2\left(-e^{x}\right)-2\text{Li}_3\left(-e^{x}\right)\right]+\mathcal{O}\left[\left(\frac{k_\text{B}T}{\gamma}\right)^3\right]\right\},
\end{split}
\end{equation}
where $x=\Delta\mu/(k_\text{B}T)$, $\text{Li}_n(\cdot)$ is the polylogarithm function and $'$ denotes the differentiation with respect to $E/\gamma$. The other integrals in the last line of Eq.~\eqref{eq:charge-shot-general} can be handled in the same way. Eventually, one finds, keeping terms up to second order in $k_\text{B}T/\gamma$,
\begin{equation}
\begin{split}
    S^I_\text{sh}&=\frac{4e^2}{h}k_\text{B}T\Bigg\{F(0)\left[\ln\left(2+2\cosh x\right)-1\right]-\frac{k_\text{B}T}{\gamma}F'(0)\left[x\ln\left(2+2\cosh x\right)+\mathcal{L}_2(x)\right]\\
    &\qquad\qquad\quad+\left(\frac{k_\text{B}T}{\gamma}\right)^2F''(0)\left[\frac{1}{2}x^2\ln\left(2+2\cosh x\right)+x\,\mathcal{L}_2(x)-\mathcal{L}_3(x)-\frac{\pi^2}{6}\right]+\mathcal{O}\left[\left(\frac{k_\text{B}T}{\gamma}\right)^3\right]\Bigg\},
\end{split}
\label{eq:charge-shot-general-expansion}
\end{equation}
where $\mathcal{L}_n(x)=\text{Li}_n(-e^x)+(-1)^{n+1}\text{Li}_n(-e^{-x})$.
In a similar way, we evaluate the thermal noise as follows:
\begin{equation}
    S^I_\text{th}=\frac{4e^2}{h}\int_{-\infty}^{+\infty}D(E/\gamma)f_\lef(E)[1-f_\lef(E)]\approx \frac{4e^2}{h}k_\text{B}T\left[D(0)+D''(0)\left(\frac{k_\text{B}T}{\gamma}\right)^2\frac{\pi^2}{6}\right]\,.
    \label{eq:charge-thermal-general-expansion}
\end{equation}
The only thing left is to find the thermovoltage $\Delta\mu_I$ to be plugged in Eq.~\eqref{eq:charge-shot-general-expansion} in order to have the shot noise at zero current. To this aim, we have to impose the condition
\begin{equation}
    I=-\frac{2e}{h}\int_{-\infty}^{+\infty}D(E/\gamma)[f_\lef(E)-f_\ri(E)]=0\,.
\end{equation}
By performing the same expansion as in the previous calculations, we find
\begin{equation}
    \Delta\mu_I^2 D'(0)-2\gamma\Delta\mu_I D(0)-\frac{\pi^2(k_\text{B}T)^2}{3}D'(0)=0\implies\Delta\mu_I=\frac{\gamma D(0)}{D'(0)}\left[1-\sqrt{1+\frac{\pi^2}{3}\left(\frac{k_\text{B}T}{\gamma}\right)^2\left(\frac{D'(0)}{D(0)}\right)^2}\right].
    \label{eq:thermovoltage-expansion}
\end{equation}
The sign of the square root is chosen in order to have the physically relevant solution, which requires that when $D'(0)\to 0$ one must recover $\Delta\mu_I\to 0$, meaning that no thermoelectric effect is present to this order of approximation. 

From Eqs.~\eqref{eq:charge-shot-general-expansion} and \eqref{eq:charge-thermal-general-expansion} we can compute an approximation for the ratio $R_I$ in terms of the thermovoltage given by Eq.~\eqref{eq:thermovoltage-expansion} and the transmission function and its derivatives evaluated at the energy $E=\mu_\lef=0$. As we show below, the leading terms in Eqs.~\eqref{eq:charge-shot-general-expansion}-\eqref{eq:charge-thermal-general-expansion} already provide a good approximation; we then find the simplified expression
\begin{equation}
    R_I=\frac{S^I_\text{sh}}{S^I_\text{th}}\bigg|_{I=0}\approx[1-D(0)]\left[\ln\left(2+2\cosh\frac{\Delta\mu_I}{k_\text{B}T}\right)-1\right]\,,
    \label{eq:ratio-approx}
\end{equation}
which is the result reported in the main text.

For a constant transmission $D(E)=D^0$ the thermovoltage is $\Delta\mu_I=0$ and all derivatives of $D(E)$ are zero. Therefore the previous expression at $\Delta\mu_I=0$ actually gives the exact result in this case:
\begin{equation}
    R_I|_{D(E)=D^0}=(1-D^0)(2\ln 2-1).
    \label{eq:ratio-constant}
\end{equation}
Similar expressions have been previously obtained in this regime~\cite{Larocque2020}.

We now illustrate the approximation \eqref{eq:ratio-approx} by considering as a specific example a Lorentzian transmission function, given by
\begin{figure}[t]
\begin{center}
\includegraphics[width=\textwidth]{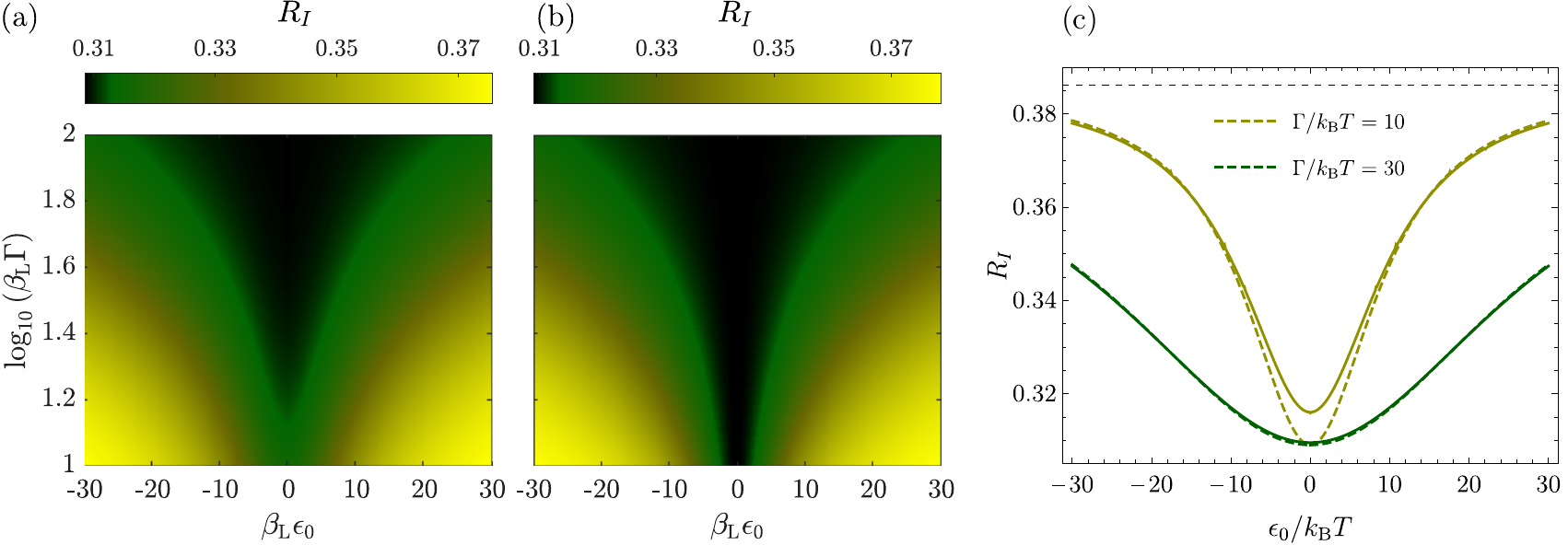}
\end{center}
\caption{(a) Exact ratio, computed by numeric integration. (b) Ratio $R_I$ calculated with the approximation. (c) Comparison between the approximation (dashed lines) and the exact result (solid lines) for two representative values of $\Gamma/k_\text{B}T$. In all the plots a Lorentzian transmission function centered at energy $\epsilon_0$ and with $D^0=0.2$ has been considered.}
\label{fig:density}
\end{figure}
\begin{equation}
    D_\text{Lor}\left(\frac{E}{\Gamma}\right)=\frac{D^0}{1+\left(\frac{E}{\Gamma}-\frac{\epsilon_0}{\Gamma}\right)^2}\,.
    \label{eq:lor-trans}
\end{equation}
In this case, the typical energy scale set by the transmission is $\Gamma$ (notice that we are using the same convention as in the main text, where $\gamma$ refers to the energy scale of a generic transmission, whereas $\Gamma$ refers to the specific choice in \eqref{eq:lor-trans}).
In Fig.~\ref{fig:density}(a-b) we compare the ratio $R_I$ given in Eq.~\eqref{eq:ratio-approx} with the exact result for the same quantity, computed by numerical integration. As we can see, the agreement between the two is very good for $k_\text{B}T/\Gamma<10^{-1}$. Fig.~\ref{fig:density}(c) shows two different cuts of the density plots in the top panel and illustrates that the approximation is increasingly good the smaller $k_\text{B}T/\Gamma$. Finally, as discussed in the main text, notice how the ratio tends to the value $2\ln 2-1$ when $\epsilon_0$ is the largest energy scale. This is because in this situation, the transmission function for the relevant energies in the transport window appears as a constant $D^0\ll 1$ (tails of the Lorentzian), in which case Eq.~\eqref{eq:ratio-constant} applies and $1-D^0\approx 1$.

\section{Heat noise for a general transmission}
Following the lines of the calculation in Sec.~\ref{sec:app:charge-large}, it is possible to obtain an approximate expression for the heat noise too. The calculation being very similar to the one presented above, here we simply report the results. We assume again the conditions $k_\text{B}T_\ri\ll k_\text{B}T_\lef\equiv k_\text{B}T\ll\gamma$ and we consider the heat noise in the left reservoir. The heat shot noise can be expressed as
\begin{equation}
    S^{J}_\text{sh}=\frac{4}{h}(k_\text{B}T)^3\left\{A(x)F(0)+\left(\frac{k_\text{B}T}{\gamma}\right)B(x)F'(0)+\left(\frac{k_\text{B}T}{\gamma}\right)^2C(x)F''(0)+\mathcal{O}\left[\left(\frac{k_\text{B}T}{\gamma}\right)^3\right]\right\},
    \label{eq:heat-shot-general-expansion}
\end{equation}
where $x=\Delta\mu/(k_\text{B}T)$ and
\begin{align}
    A(x)&=-\frac{\pi^2+x^3}{3}+2x^2\ln(1+e^{x})+4x\text{Li}_2(-e^{x})-4\text{Li}_3(-e^{x})\,,\\
    B(x)&=\frac{x^4}{4}-\frac{7\pi^4}{60}-2x^3\ln(1+e^x)-6x^2\text{Li}_2(-e^x)+12x\text{Li}_3(-e^x)-12\text{Li}_4(-e^x)\,,\\
    C(x)&=-\frac{7\pi^4+3x^5}{30}+x^4\ln(1+e^x)+4x^3\text{Li}_2(-e^x)-12x^2\text{Li}_3(-e^x)+24x\text{Li}_4(-e^x)-24\text{Li}_5(-e^x)\,.
\end{align}
Similarly, for the thermal heat noise the result is
\begin{equation}
    S_\text{th}^J=\frac{4}{h}(k_\text{B}T)^3\frac{\pi^2}{3}\left[D(0)+\left(\frac{k_\text{B}T}{\gamma}\right)^2\frac{7\pi^2}{10}D''(0)\right]\,.
    \label{eq:heat-thermal-general-expansion}
\end{equation}
By imposing that the heat current $J_\lef$ in the left reservoir be zero, one finds the value of the (dimensionless) heat thermovoltage $x_J=\Delta\mu_J/(k_\text{B}T)$. Keeping terms up to the first order in $(k_\text{B}T/\gamma)$, we find a cubic equation for $x_J$, with the three real solutions $(k=0,1,2)$
\begin{equation}
    x_J^{(k)}=\frac{\gamma}{k_\text{B}T}\frac{D(0)}{2D'(0)}+\frac{\gamma}{k_\text{B}T}\left|\frac{D(0)}{D'(0)}\right|\cos\left(\frac{1}{3}\arccos\left\{\text{sgn}(D'(0))\left[1-2\pi^2\left(\frac{k_\text{B}T}{\gamma}\right)^2\left(\frac{D'(0)}{D(0)}\right)^2\right]\right\}-\frac{2\pi k}{3}\right)\,.
    \label{eq:xj}
\end{equation}
Only two of them are physically relevant, namely $x_J^{(1)}$ and $x_J^{(2)}$ when $D'(0)>0$ and $x_J^{(0)}$ and $x_J^{(1)}$ when $D'(0)<0$. Consistently with the approximation $k_\text{B}T\ll\gamma$, the previous equation may be simplified, yielding
\begin{equation}
    x_J=\pm\frac{\pi}{\sqrt{3}}\left[1\pm\frac{\pi}{3\sqrt{3}}\frac{k_\text{B}T_\lef}{\gamma}\frac{D'(0)}{D(0)}\right]\,.
    \label{eq:xj_simplified}
\end{equation}
The leading order approximation to the ratio between thermal and shot noise then reads 
\begin{equation}
    R_J=\frac{S_\text{sh}^{J}}{S_\text{th}^J}\bigg|_{J=0}\approx\frac{3}{\pi^2}[1-D(0)]A(x_J)\,,
\end{equation}
which is the result in the main text.

For a constant transmission $D(E)=D^0$, all the derivatives vanish and therefore the first terms in Eqs.~\eqref{eq:heat-shot-general-expansion} and \eqref{eq:heat-thermal-general-expansion} give the exact results for the heat noise (shot and thermal, respectively) at arbitrary voltage. Moreover, the solutions for the heat thermovoltage are $x_J=\pm\pi/\sqrt{3}$, which can be obtained from Eq.~\eqref{eq:xj} or ~\eqref{eq:xj_simplified} by taking the limit $D'(0)\to 0$. As for the charge noise, one finally finds the exact expression shown in the main text.
\begin{equation}
    R_J|_{D(E)=D^0}=(1-D^0)\frac{3}{\pi^2}A\left(\frac{\pi}{\sqrt{3}}\right)=R_J^0(1-D^0)\approx 0.4469(1-D^0)\,.
    \label{eq:app:heat-bound}
\end{equation}

\section{Noise ratio at arbitrary temperatures for constant transmission}
\begin{figure}[t]
\begin{center}
\includegraphics[width=0.8\textwidth]{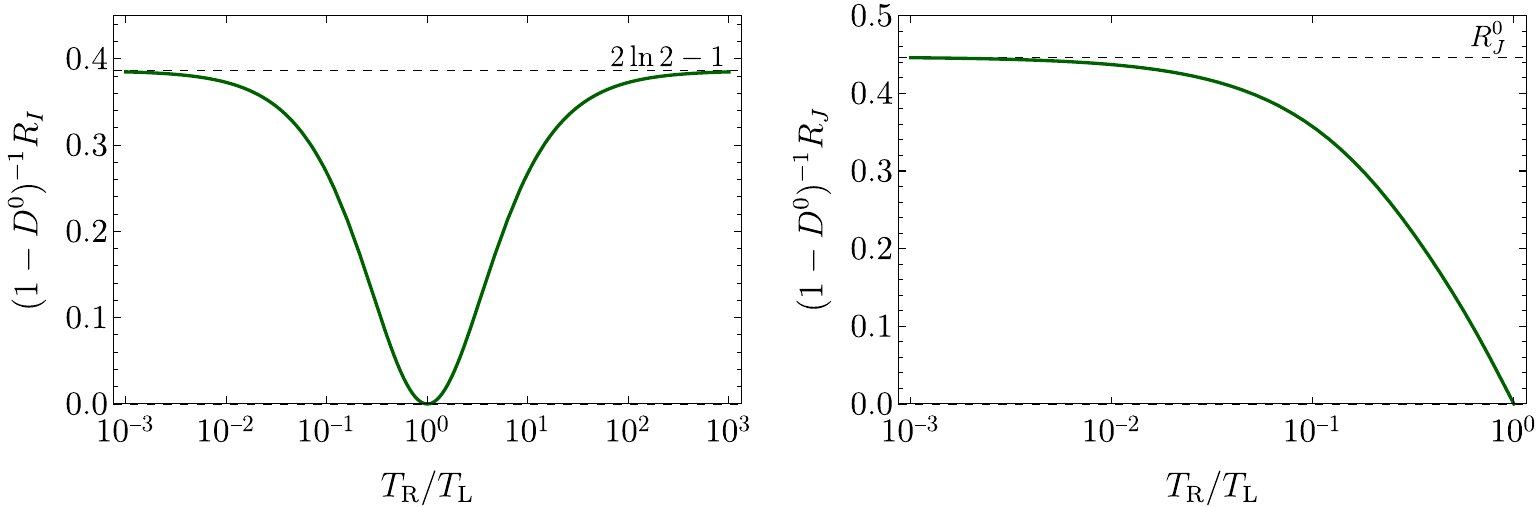}
\end{center}
\caption{Noise ratios $R_I$ and $R_J$ at arbitrary temperatures for a constant transmission function $D(E)=D^0$. These plots are obtained with a numerical evaluation of the second integral in Eq.~(2) of the main text, with a voltage bias such that the charge or the heat current vanishes. The dashed lines show the position of the upper bounds.}
\label{fig:supp:ratios-const}
\end{figure}
In this section, we finally show that the exact results in Eqs.~\eqref{eq:ratio-constant} and \eqref{eq:app:heat-bound} are actually upper limits for the charge and heat noise ratios $R_I$ and $R_J$, when the transmission probability does not depend on energy, namely $D(E)=D^0$. The validity of this statement is shown in Fig.~\ref{fig:supp:ratios-const}, where we show the ratios $R_I$ and $R_J$ for an arbitrary choice of the temperatures $T_\text{R}$ and $T_\text{L}$. We clearly see that the $R_I$ approaches the upper limit $(1-D^0)(2\ln 2-1)$ when either $T_\text{R}\ll T_\text{L}$, as considered in the main text, or $T_\text{R}\gg T_\text{L}$. For all intermediate temperatures, $R_I$ is always smaller than this value and it vanishes when $T_\text{R}=T_\text{L}$ because the zero-current shot noise is zero under this condition.

Concerning the heat noise ratio $R_J$, we observe the same behavior: it approaches the upper limit $R_J^0$ when $T_\text{R}\ll T_\text{L}$ and it decreases when $T_\text{R}$ increases. As a difference compared to the charge noise, notice that the plot for $R_J$ is only shown for $T_\text{R}\le T_\text{L}$. This is because the heat current into the left reservoir never vanishes when $T_\text{R}>T_\text{L}$.

\section{Charge and heat currents for a Lorentzian transmission}
In this section, we present analytical expressions for charge and heat currents for a conductor with an arbitrarily sharp Lorentzian-shaped transmission probability. Starting from Eqs.~(1) and (3) in the main text, we find for the charge current at any temperature and voltage bias
\begin{equation}
    I=-\frac{eD^0\Gamma}{h}\,\,\text{Im}\!\sum_{\alpha=\ri,\lef}\vartheta_\alpha\left[\Psi_\alpha(z)-2\pi if_\alpha(-iz)\right]\,.
\end{equation}
Here, $\Psi_\alpha(x)=\psi(\frac{1}{2}+\frac{\beta_\alpha(z-i\mu_\alpha)}{2\pi})-\psi(\frac{1}{2}-\frac{\beta_\alpha(z-i\mu_\alpha)^*}{2\pi})$, with $\psi$ the digamma function, $\vartheta_{\ri/\lef}=\mp1$, and $z=\Gamma+i\epsilon_0$.

For the heat current in contact L, we find for arbitrary $\Delta T$ and $\Delta\mu$
\begin{equation}
    J_\lef=\frac{D^0\Gamma}{h}\bigg\{2\Gamma\ln\frac{T_\lef}{T_\ri}+\,\text{Re}\!\!\sum_{\alpha=\ri,\lef}\!z\vartheta_\alpha\left[\tilde\Psi_\alpha(z)-2\pi i f_\alpha(-iz)\right]\bigg\},
\end{equation}
with $\tilde\Psi_\alpha(z)=\psi(\frac{1}{2}+\frac{\beta_\alpha(z-i\mu_\alpha)}{2\pi})+\psi(\frac{1}{2}-\frac{\beta_\alpha(z-i\mu_\alpha)}{2\pi})$. 

These analytic expressions are the ideal starting point for the evaluation of the charge and heat thermovoltages, as they substantially simplify the finding of numerical solutions for $I\equiv0$ or $J_\lef\equiv0$.

\section{Additional plots for noise ratios with a Lorentzian transmission}
\begin{figure}[t]
\begin{center}
\includegraphics[width=0.7\textwidth]{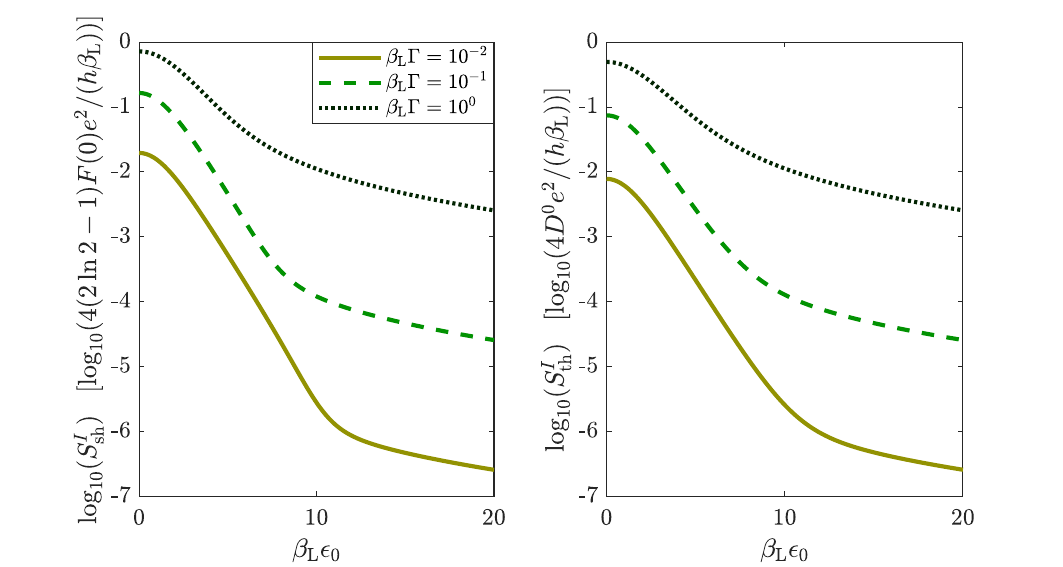}
\end{center}
\caption{Zero-current charge thermal and shot noise, $S_\text{th}^I$ and $S_\text{sh}^I$, as a function of $\epsilon_0$ and for different values of the Lorentzian width $\Gamma$. The plots refer to the large temperature bias scenario, $T_\ri/T_\lef=0$. Noises are normalized to the value they take for a constant transmission, $S^I_\text{th}|_{D(E)=D^0}$ and $S^I_\text{sh}|_{D(E)=D^0}$. The plots show that the reduction at $\epsilon_0=0$ is more relevant for the thermal contribution and this leads to an increased ratio $R_I$ exceeding the constant-transmission bound in Eq.~\eqref{eq:ratio-constant}.}
\label{fig:supp:charge-separate}
\end{figure}

In this section, we show additional data concerning specific regimes which have been only briefly mentioned in the main text. We start by analyzing in more detail the charge noise ratio $R_I$ when $\epsilon_0\to 0$. In this case, due to particle-hole symmetry, the thermovoltage is $\Delta\mu_I=0$ and therefore only a temperature bias is present, leading to a purely delta-$T$ noise, as in the case of a constant transmission. Still, we have shown in the main text that the energy-dependence of $D(E)$ results in a ratio $R_I$ exceeding the constant-transmission value in Eq.~\eqref{eq:ratio-constant}. In Fig.~\ref{fig:supp:charge-separate} we show that this happens as, while both the thermal and shot contributions are reduced compared to the constant transmission case, the reduction is more pronounced for $S^I_\text{th}$, resulting in an increased ratio $R_I$ when $\epsilon_0\to 0$, as observed in Fig.~2(c) in the main text. {To show this effect more clearly, in Fig.~\ref{fig:supp:charge-separate} we normalized each noise contribution to the value it takes at constant transmission, namely $S_\text{th}^I|_{D(E)=D^0}=4e^2D^0/(h\beta_\lef)$ and $S_\text{sh}^I|_{D(E))D^0}=4e^2D^0(1-D^0)/(h\beta_\lef)(2\ln 2-1)$.}
\begin{figure}[t]
\begin{center}
\includegraphics[width=0.9\textwidth]{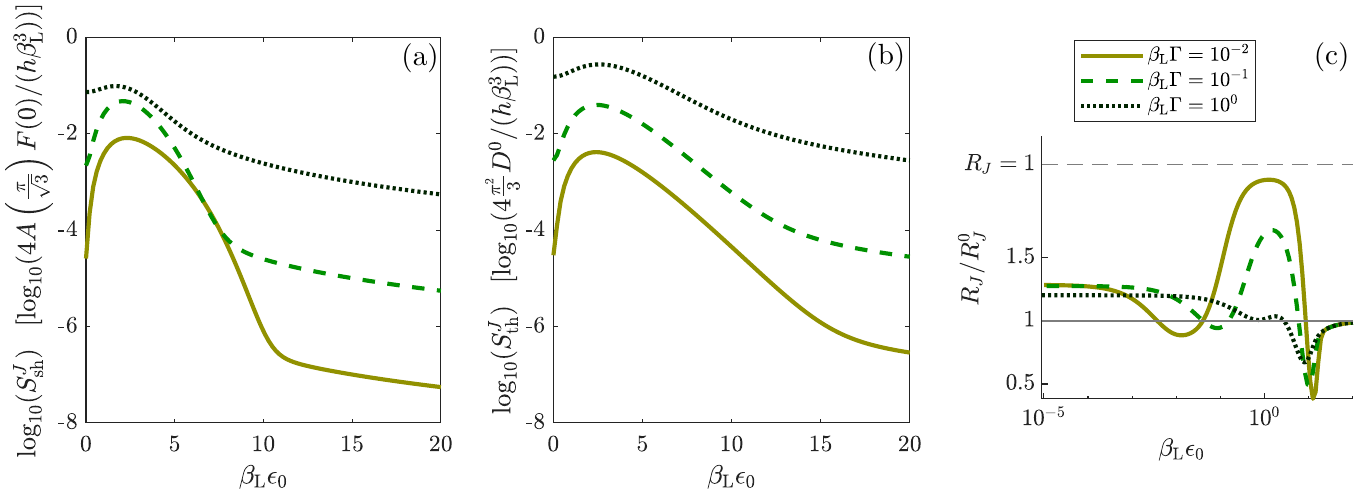}
\end{center}
\caption{(a) Zero-current thermal and shot heat noise, $S_\text{th}^J|_{J=0}$ and $S_\text{sh}^J|_{J=0}$, as functions of $\epsilon_0$ for different values of $\Gamma$. The two types of noise are normalized to the value they take at constant transmission. The thermal heat noise decreases more rapidly compared to the shot heat noise when $\Gamma$ is decreased. This results in regions where the heat noise ratio is greater than in the case of constant transmission. (b) Heat noise ratio as a function of $\epsilon_0$, normalized to the value at constant transmission. The region $10^{-5}<\beta_\lef\epsilon_0<10^{-1}$ shows a non-trivial behaviour, which is not visible with a linear scale for $\epsilon_0$.}
\label{fig:supp:heat-separate}
\end{figure}

Next, we consider the same particle-hole symmetric situation in the case of the heat noise. Here, a similarly increased ratio $R_J$ was observed [Fig.~3(c) in the main text], but its behavior as $\epsilon_0\to 0$ is more complicated than that of $R_I$. This is because, while both $S_\text{th}^I$ and $S_\text{sh}^I$ are monotonically increasing when $\epsilon_0\to 0$, their heat counterparts exhibit a maximum at a finite $\epsilon_0$, after which they decrease as $\epsilon_0\to 0$. This behavior is shown in Fig.~\ref{fig:supp:heat-separate} The way $S_\text{th}^J$ and $S_\text{sh}^J$ decrease as $\epsilon\to 0$ depends on $\Gamma$ and thus different behaviors for the ratio $R_J$ are observed when $\Gamma$ is varied.